\documentclass[11pt,superscriptaddress,preprintnumbers,amsmath,amssymb,nofootinbib]{revtex4-1}
\pdfoutput=1
\usepackage{graphicx}
\usepackage{epstopdf}
\usepackage{dcolumn}
\usepackage{bm}
\usepackage{hyperref}
\usepackage{color}
\usepackage{amsmath}
\usepackage{slashed}
\usepackage{cancel}
\usepackage{xpatch}

\begin{document}


\def\a{\alpha}
\def\b{\beta}
\def\c{\varepsilon}
\def\d{\delta}
\def\e{\epsilon}
\def\f{\phi}
\def\g{\gamma}
\def\h{\theta}
\def\k{\kappa}
\def\l{\lambda}
\def\m{\mu}
\def\n{\nu}
\def\p{\psi}
\def\q{\partial}
\def\r{\rho}
\def\s{\sigma}
\def\t{\tau}
\def\u{\upsilon}
\def\v{\varphi}
\def\w{\omega}
\def\x{\xi}
\def\y{\eta}
\def\z{\zeta}
\def\D{\Delta}
\def\G{\Gamma}
\def\H{\Theta}
\def\L{\Lambda}
\def\F{\Phi}
\def\P{\Psi}
\def\S{\Sigma}

\def\o{\over}
\def\beq{\begin{align}}
\def\eeq{\end{align}}
\newcommand{\gsim}{ \mathop{}_{\textstyle \sim}^{\textstyle >} }
\newcommand{\lsim}{ \mathop{}_{\textstyle \sim}^{\textstyle <} }
\newcommand{\vev}[1]{ \left\langle {#1} \right\rangle }
\newcommand{\bra}[1]{ \langle {#1} | }
\newcommand{\ket}[1]{ | {#1} \rangle }
\newcommand{\EV}{ {\rm eV} }
\newcommand{\KEV}{ {\rm keV} }
\newcommand{\MEV}{ {\rm MeV} }
\newcommand{\GEV}{ {\rm GeV} }
\newcommand{\TEV}{ {\rm TeV} }
\newcommand{\1}{\mbox{1}\hspace{-0.25em}\mbox{l}}
\newcommand{\headline}[1]{\noindent{\bf #1}}
\def\diag{\mathop{\rm diag}\nolimits}
\def\Spin{\mathop{\rm Spin}}
\def\SO{\mathop{\rm SO}}
\def\O{\mathop{\rm O}}
\def\SU{\mathop{\rm SU}}
\def\U{\mathop{\rm U}}
\def\Sp{\mathop{\rm Sp}}
\def\SL{\mathop{\rm SL}}
\def\tr{\mathop{\rm tr}}
\def\mpl{M_{\rm Pl}}

\def\IJMP{Int.~J.~Mod.~Phys. }
\def\MPL{Mod.~Phys.~Lett. }
\def\NP{Nucl.~Phys. }
\def\PL{Phys.~Lett. }
\def\PR{Phys.~Rev. }
\def\PRL{Phys.~Rev.~Lett. }
\def\PTP{Prog.~Theor.~Phys. }
\def\ZP{Z.~Phys. }

\def\dd{\mathrm{d}}
\def\ff{\mathrm{f}}
\def\BH{{\rm BH}}
\def\inf{{\rm inf}}
\def\ev{{\rm evap}}
\def\eq{{\rm eq}}
\def\SM{{\rm sm}}
\def\Mpl{M_{\rm Pl}}
\def\GeV{{\rm GeV}}
\newcommand{\Red}[1]{\textcolor{red}{#1}}
\newcommand{\TL}[1]{\textcolor{blue}{\bf TL: #1}}

\title{
Naturalness and the muon magnetic moment
}

\author{Nima Arkani-Hamed}
\author{Keisuke Harigaya}
\address{School of Natural Sciences, Institute for Advanced Study, Princeton, NJ 08540, USA}

\begin{abstract}
We study a predictive model for explaining the apparent deviation of the muon anomalous magnetic moment from the Standard Model expectation. There are no new scalars and hence no new hierarchy puzzles beyond those associated with the Higgs; the only new particles at the TeV scale are vector-like singlet and doublet leptons. Interestingly, this simple model provides a calculable example violating the Wilsonian notion of naturalness:~despite the absence of any symmetries prohibiting its generation, the coefficient of the naively leading dimension-six operator for $(g-2)$ vanishes at one-loop. While effective field theorists interpret this either as a  surprising UV cancellation of power divergences, or as a delicate cancellation between matching UV and calculable IR corrections to $(g-2)$ from parametrically separated scales, there is a simple explanation in the full theory:~the loop integrand is a total derivative of a function vanishing in both the deep UV and IR. The leading contribution to $(g-2)$ arises from dimension-eight operators, and thus the required masses of new fermions are lower than naively expected, with a sizeable portion of parameter space already covered by direct searches at the LHC.
The viable parameter space free of fine-tuning for the muon mass will be fully covered by future direct LHC searches, and {\it all} of the parameter space can be probed by precision measurements at planned future lepton colliders. 
\end{abstract}

\date{\today}

\maketitle

\section{Introduction}
Precise measurements of the muon anomalous magnetic moment, $(g-2)_\mu$, are sensitive to the interactions of the muon with
new particles.
The BNL E821 experiment~\cite{Bennett:2006fi} observed a deviation of $(g-2)_\mu$ from the Standard Model (SM) prediction as large as the electroweak contribution. The SM prediction has been improved~\cite{Aoyama:2020ynm} and the BNL measurement has been confirmed by the Fermilab E989 experiment~\cite{Abi:2021gix}. The deviation is now
\begin{align}
    \Delta a_\mu = \frac{(g-2)_{\rm exp} - (g-2)_{\rm SM}}{2} = (2.51 \pm 0.59)\times 10^{-9}.
\end{align}
While the status of the SM computation remains uncertain--given the tension between the data-driven approaches to hadronic vacuum polarization contributions going into the above deviation and recent high-precision lattice QCD simulations that appear to largely eliminate the anomaly~\cite{Borsanyi:2020mff}--for the purposes of this note we will take this measurement as a hint of new physics.

Since $(g-2)_{\mu}$ breaks the chiral symmetry of the muon, quantum corrections generating $(g-2)_{\mu}$, with the external photon removed, necessarily generate a muon mass, $\Delta m_\mu$. Assuming that $\Delta m_\mu$ and $(g-2)_{\mu}$ are given by the dimension-four and six operators respectively, the relation between $\Delta a_\mu$, $\Delta m_\mu$, and the new physics mass scale $M$ is generically given by
\begin{align}
    \frac{\Delta a_\mu}{m_\mu} \sim \frac{\Delta m_\mu}{M^2}.
\end{align}
If $\Delta m_\mu$ is a perturbative correction to the tree-level muon Yukawa coupling, $\Delta m_\mu \sim \alpha_2 m_\mu/(4\pi)$, the expected new particle masses are around $M \sim 100$ GeV. Charged particles near 100 GeV would have been copiously produced at the Large Hadron Collider (LHC), which has not seen any such signals. Thus one needs an enhancement of $\Delta a_\mu$ in comparison with $\Delta m_\mu$, such as the ${\rm tan} \beta$ enhancement in supersymmetric theories~\cite{Lopez:1993vi,Chattopadhyay:1995ae,Moroi:1995yh}.
If, on the other hand, $\Delta m_\mu$ originates from muon chiral symmetry breaking beyond the muon Yukawa coupling, it may be as large as $m_\mu$ (or larger if the muon mass is fine-tuned) without loop suppression~\cite{Borzumati:1999sp,Czarnecki:2001pv}. Then the expected new physics scale is $M\sim 2$ TeV. The present LHC constraints are easily satisfied, but the new particles may be beyond the reach of the LHC and near-future colliders.

There are a priori a huge range of possibilities for the new particles running inside the loop responsible for $(g-2)_\mu$, many of which have been explored for several decades. But the larger theoretical context in which to consider possible new physics explanations for $(g-2)_\mu$ has changed radically over the past decade, due to the absence of ``natural new physics" at the LHC to explain the origin of the Higgs mass scale. This suggests at least a ``little hierarchy" between the weak scale and the cutoff of the SM effective theory, and makes it even more plausible than it may already have been in the past, to imagine that the Higgs is tuned to be light for anthropic reasons. If we take this picture seriously, we are led to a much more constrained set of possibilities for explaining $(g-2)_\mu$, since there is no reason to have any {\it other} light scalars at the TeV scale--as they serve no anthropic purpose--and absent light scalars to higgs them, we should not expect new gauge bosons either. Thus we can only imagine theories with new vector-like fermions at the TeV scale, whose masses are protected by chiral symmetries. 

Models for $(g-2)_\mu$ motivated by this philosophy were investigated in~\cite{Kannike:2011ng,Dermisek:2013gta}. Emphasis was put on models with electromagnetic charged vector-like leptons in $SU(2)_L$ singlets or triplets.
In order to generate a large enough $(g-2)_\mu$, in these theories the significant portion of the muon mass arises from the tree-level exchange of heavy vector-like leptons generating a dimension-six operator $\ell e^c H |H|^2$,
and it is expected that the $h\mathchar`-\mu\bar{\mu}$ coupling
deviates from the SM prediction. 
The recent upper bound on the coupling~\cite{Aad:2020xfq,Sirunyan:2020two} has excluded some of the parameter space. Interestingly, in the model with an $SU(2)$ doublet and an hyper-charged $SU(2)$ singlet, the central value of the $(g-2)_\mu$ anomaly is such that in a good portion of parameter space, the $h\mathchar`-\mu \bar{\mu}$ coupling is $\sim (-1) \times$ that in the SM, so the rate for $h\rightarrow \mu \bar{\mu}$ is unaffected.

Perhaps the simplest possible model along these lines has a single new vector-like $SU(2)_L$ doublet and a gauge singlet. 
The tree-level exchange of the heavy fermions does not give a muon Yukawa coupling, so that the $h\mathchar`-\mu\bar{\mu}$ coupling is guaranteed to be the SM-like.
This setup was also briefly mentioned in~\cite{Kannike:2011ng} and analyzed in~\cite{Freitas:2014pua}. 

In this note, we would look to draw attention to a very simple but intriguing feature of $(g-2)_\mu$ in this model, which also has immediate phenomenological implications. We find that ironically, this model, whose sparse structure was motivated by the ``unnaturalness" of the SM Higgs, itself gives sharp violation of Wilsonian naturalness in the computation of $(g-2)_\mu$! As we will see, despite the absence of any obvious symmetry that prevents the generation of $(g-2)_\mu$ by a dimension-six operator, it vanishes at one-loop level. In the effective field theory after integrating out
both vector-like fermions, the vanishing $(g-2)_\mu$ is seen as a surprising absence of quadratically (or higher power) divergent correction to it. In the effective field theory
at energy scales between the two vector-like fermions,
the vanishing $(g-2)_\mu$ is seen as a  delicately fine-tuned cancellation between
a calculable IR contribution to $(g-2)_\mu$
and a matching UV contribution.
In the full UV theory,
all of this is understood as a consequence of the fact that the loop integrand is a total derivative of a function that vanishes, for simple reasons, in both the deep UV and deep IR.

Apart from giving a concrete examples of ``absence of power divergences" and ``UV-IR correlation mechanism" at work violating the Wilsonian naturalness, this phenomenon has an important phenomenological consequence. The actual leading contribution to  $(g-2)_\mu$ comes from a dimension-eight operator whose effect is suppressed by $m_W^2/M^2$, and hence the masses of the new fermions are  required to be smaller that the $M \sim 2$ TeV scale expected from the naive estimates to explain the $(g-2)_\mu$ anomaly.  As result, a sizeable portion of the parameter space has been already ruled-out by direct searches for vector-like leptons at the LHC.
The portion of the remaining parameter space that is free of fine-tuning for the muon mass will be incisively probed by future direct LHC searches, while {\it all}  of the parameter space can be probed by precision measurements at planned future lepton colliders.

\section{The model and anomalous magnetic moment}

We extend the SM by vector-like leptons $L$, $L^c$, $S$, and $S^c$. 
$L/L^c$ have the same/opposite gauge charge as/to the SM lepton doublets and $SS^c$ are gauge singlets.
We introduce the following Yukawa couplings and Dirac masses,
\begin{align}
\label{eq:L}
 {\cal L } = & - Y_L \ell S^c H^\dag - Y_R L e^c H - Y_V S L^c  H - Y_V' L S^c H^\dag  - m_L L L^c - m_S S S^c  + {\rm h.c.},
\end{align}
where $H$ is the SM Higgs, $\ell$ is the doublet containing the left-handed muon, and $e^c$ is the right-handed muon.

Since $\ell$ couples only to a neutral fermion $S^c$, the tree-level exchange of the heavy fermions does not generate a muon Yukawa coupling.
One-loop quantum corrections
given by Fig.~\ref{fig:dim6} without the external photon lines generate a muon Yukawa coupling,
\begin{align}
\label{eq:mass}
    \Delta y_\mu =&  -\frac{Y_L Y_R }{16\pi^2} \left(Y_V\frac{m_S}{m_L} + Y_V' \right)   \left( \frac{m_S^2}{m_S^2 - m_L^2}{\rm log}\frac{\Lambda^2}{m_S^2} + \frac{m_L^2}{m_L^2 - m_S^2}{\rm log}\frac{\Lambda^2}{m_L^2}  \right),
\end{align}
where $\Lambda$ is the cut-off scale of the theory. In our summary discussion, we describe the embedding of the setup into a ``minimally split" supersymmetric theory with scalar masses around $100-1000$ TeV scales, where the muon mass (together with the electron and tau masses) may be fully radiatively generated, but now turn to the computation of $(g-2)_\mu$.

\subsection{One-loop correction to dimension-six operator}

\begin{figure}[t]
\begin{center}
 \includegraphics[width=0.4\textwidth]{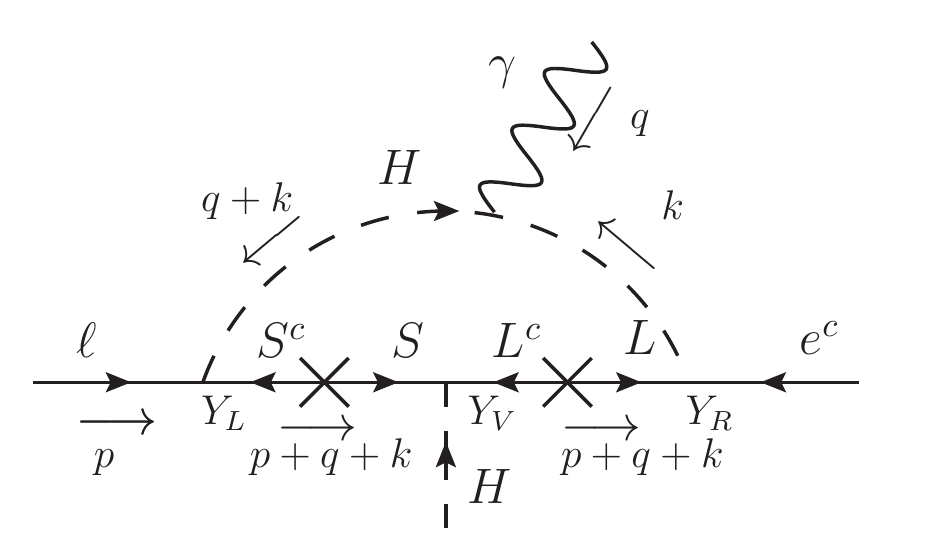}
  \includegraphics[width=0.4\textwidth]{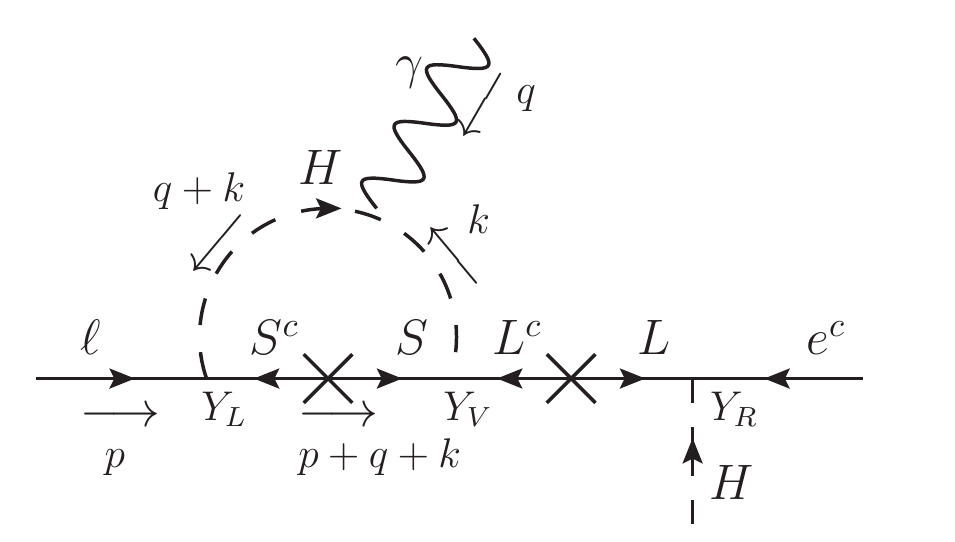}
   \includegraphics[width=0.34\textwidth]{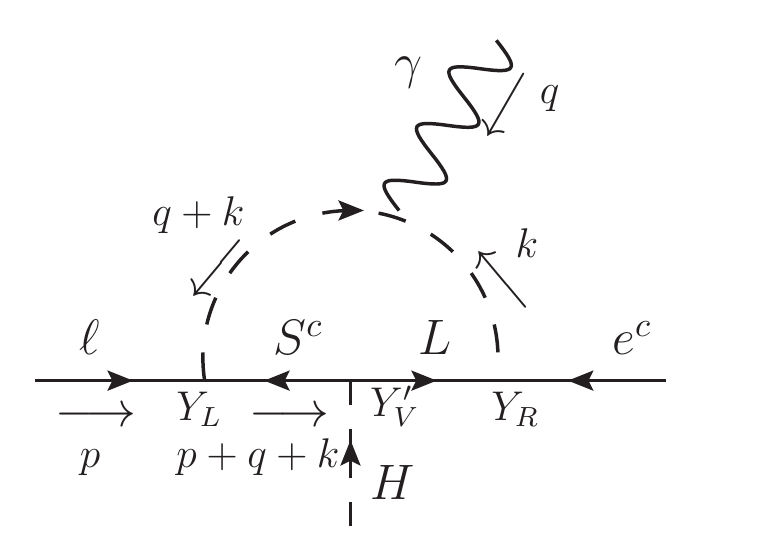}
 \caption{
 Leading diagrams for $(g-2)_\mu$. While these diagrams naively generate dimension-six operators, the bottom diagram, and the sum of the top two diagrams, actually give vanishing leading contribution to $(g-2)_\mu$. 
}
\label{fig:dim6}
\end{center}
\end{figure}

At the dimension-six level, $(g-2)_\mu$ arises from left-right mixing operators $H \ell D^2 e^c$ and $H \ell \sigma^{\mu\nu} e^c F_{\mu \nu}$, or chirality conserving operators $\overline{\ell} \bar{\sigma}^\mu D^\nu \ell F_{\mu \nu}$ and $\overline{e^c} \bar{\sigma}^\mu D^\nu e^c F_{\mu \nu}$. The contribution of latter operators to $(g-2)_\mu$ is suppressed by $m_\mu^2/M^2$, and as we have discussed, require $M \sim 100$ GeV for perturbative couplings and so are excluded by direct LHC searches. We hence focus on the direct left-right mixing operators.
We naively expect that the dominant contribution comes from the correction around the energy scale $m_S, m_L \gg m_W$, so we use the Higgs picture and consider the diagrams in Fig.~\ref{fig:dim6} that would generate a dimension-six operator $H \ell D^2 e^c$. The contribution of the diagrams to $(g-2)_\mu$, however, vanishes as we show below.

It is easy to see that the  contribution to $(g-2)_\mu$ from $\epsilon^\mu q^\nu \sigma_{\mu \nu}$  is absent; putting $p ^\mu\to 0$ the only linear dependence on $q^\mu$ comes through the photon vertex and vanishes since $\epsilon \cdot q = 0$.  We may thus put $q=0$ and see the dependence on $p$ in order to compute the correction to $(g-2)_\mu$.
The correction from the bottom diagram is proportional to
\begin{align}
\label{eq:integral}
   \int \frac{d^4 k}{(2\pi)^4} \frac{\epsilon \cdot k}{(k^2)^2} f\left(\left(k+p\right)^2\right),~~f(u) = \frac{u}{(u+m_L^2)(u+m_S^2)},
\end{align}
where we have performed the Wick rotation.
The correction from the top two diagrams is proportional to
\begin{align}
\label{eq:integral2}
     \int \frac{d^4 k}{(2\pi)^4} \frac{\epsilon \cdot k}{(k^2)^2} \left( \frac{1}{\left( \left(k+p\right)^2 +m_S^2\right)\left( \left(k+p\right)^2 + m_L^2\right)} - \frac{1}{\left( \left(k+p\right)^2+m_S^2\right)\left(p^2 + m_L^2\right)} \right).
\end{align}
To compute $(g-2)_\mu$, it is enough to see the terms linear in $p$, so one may put $p^2=0$ in the $L$ propagator in the second term. Because of the partial cancellation between the two terms, the correction to $(g-2)_\mu$ from the top two diagrams is also proportional to Eq.~(\ref{eq:integral}).

Expanding the integrand in Eq.~(\ref{eq:integral}) to the linear order in $p$, the correction to $(g-2)_\mu$ is proportional to the integral
\begin{align}
\label{eq:int}
 \int_0^\infty \frac{du}{u} u f'(u),
\end{align}
where $u=k^2$. The integral of the total derivative $f^\prime(u)$ vanishes since $f(\infty)= f(0)=0$. The fact that $f(\infty) \to 0$ is an obvious consequence of dimensional analysis/UV calculability of $(g-2)_\mu$, while the fact that $f(0) \to 0$ is slightly more interesting. Mechanically, the explicit factor of $u$ in the numerator of $f(u)$ straightforwardly appears in the bottom diagram of Fig.~\ref{fig:dim6}, while it arises from a partial cancellation between the top two diagrams. In both cases there is a simple reason why we must have that $f(0)=0$: upon integrating out the massive leptons at tree-level and working at zero external momentum,  the operator with no derivatives $(\ell H^\dag) HH e^c$ identically vanishes due to anti-symmetric contraction of the $SU(2)$ indices in $HH$.

\begin{figure}[t]
\begin{center}
 \includegraphics[width=0.5\textwidth]{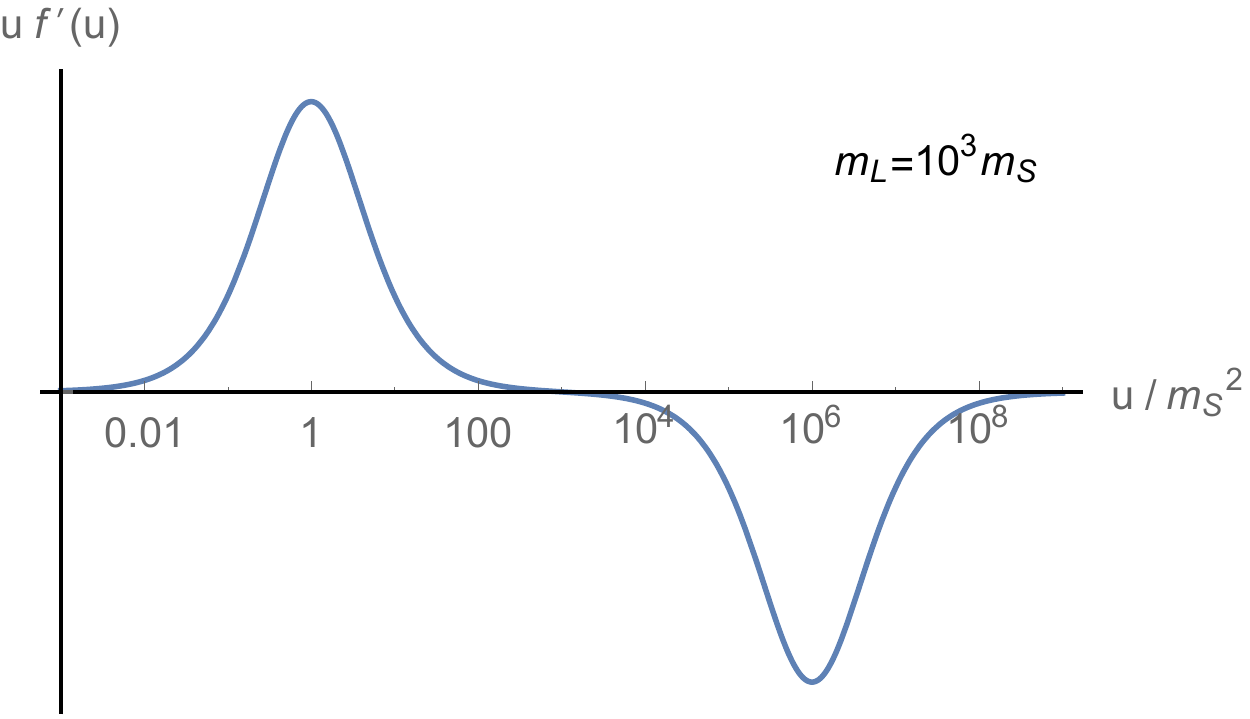}
 \caption{
 The leading order loop integrand for $(g-2)_\mu$. The integral vanishes due to a cancellation between opposite sign contributions around the scales $m_L,m_S$ of the heavy fermions, as guaranteed by the fact that the integrand is a total derivatve vanising at both zero and infinity. To clarify the structure of the integrand, we take $m_L = 10^3 m_S$.
}
\label{fig:integrand}
\end{center}
\end{figure} 

 We have understood why the full one-loop integral for $(g-2)_\mu$ is obviously zero, being a total derivative vanishing in both the deep IR and UV,  but it is also instructive to look at the energy dependence of the loop integrand to see how this happens more explicitly. The loop integrand is plotted in Fig.~\ref{fig:integrand}, where to better illustrate the point, $m_L$ is taken to be much larger than $m_S$.  We see that there are two contributions at widely different scales, near $m_S$ and $m_L$,  which ``conspire" to cancel exactly. Analytically, we can see this exact cancellation between $S$ and $L$ contributions by partial fractioning $(m_L^2 - m_S^2) f(u) = \frac{m_L^2}{u+m_L^2} - \frac{m_S^2}{u+m_S^2}$, finding for the integral
 \begin{eqnarray}
 \frac{1}{m_L^2 - m_S^2} \int \frac{du}{u} \left(\frac{\left(u/m_S^2\right)}{\left(\left(u/m_S^2\right) + 1\right)^2}- \frac{\left(u/m_L^2\right)}{\left(\left(u/m_L^2\right) + 1\right)^2} \right) \nonumber \\ = \frac{1}{m_L^2 - m_S^2} \int \left(\frac{d u_S} {(u_S + 1)^2} - \frac{d u_L}{(u_L + 1)^2} \right) = 0
 \end{eqnarray}
where $u_S \equiv (u/m_S^2)$ and $u_L \equiv (u/m_L^2)$. 

We have seen why the full integral is guaranteed to vanish. However, any effective field theory calculation chops this integral into an ``IR contribution" and a ``UV matching" part, and in any such separation, the final result of zero will appear to come from a delicate cancellation between the ``IR" and ``UV" contributions.   

\begin{figure}[t]
\begin{center}
 \includegraphics[width=0.2\textwidth]{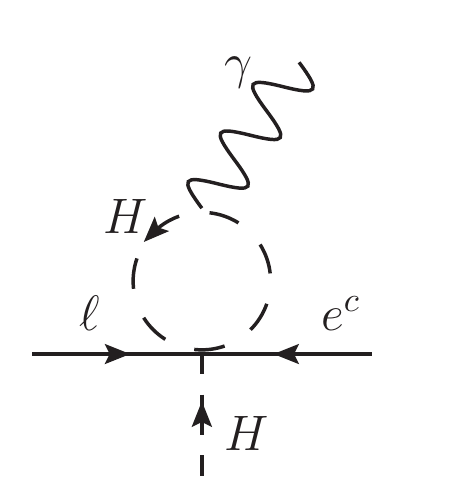}
   \includegraphics[width=0.44\textwidth]{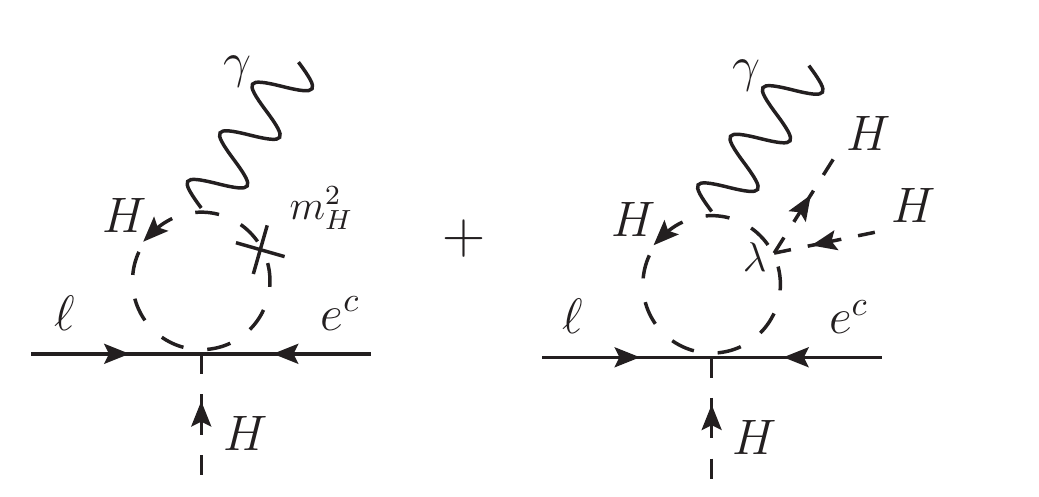}
   \includegraphics[width=0.3\textwidth]{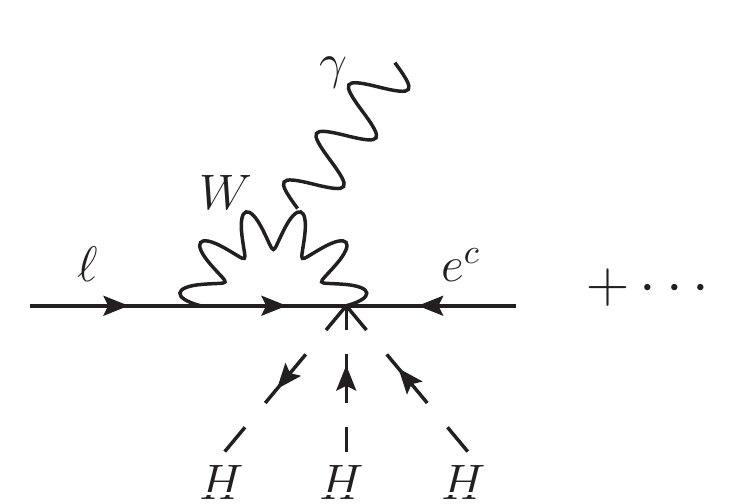}
 \caption{
 Corrections to $(g-2)_\mu$ from the dimension-eight operator in Eq.~(\ref{eq:LSout}) in the low energy effective theory after integrating out $L$ and $S$.
 ({\bf Left}) Naive quadratically divergent contribution to a dimension-six operator, which vanishes exactly in the full UV theory. ({\bf Middle}) Corrections to dimension-six and eight operators, whose sum vanishes at the vacuum. ({\bf Right}) Corrections to a dimension-eight operator that gives the dominant contribution.
}
\label{fig:quaddiv}
\end{center}
\end{figure}

Consider very low-energy effective field theorists, living at energies much smaller than both $m_S,m_L$.  Looking at the  low-energy part of the integrand beneath $m_S$ and $m_L$,  they would estimate the integral, cutting it off around the scale $m_S$. Note for small $u$, we have that $f(u) \to u/(m_S^2 m_L^2)$, thus $f^\prime(u)$ is a constant and so the integral is quadratically divergent in the UV.  This power-UV divergence has an obvious interpretation. Integrating out $S$ and $L$ at tree-level gives us a dimension-eight operator,
\begin{align}
\label{eq:LSout}
    {\cal L}_{\rm eff} = \frac{Y_L Y_R }{m_S^2 m_L^2}( Y_V \frac{m_S}{m_L} + Y_V') (\ell H^\dag) H \slashed{D}^2 (H e^c).
\end{align}
This operator breaks the chiral symmetry on the muons that protects both the dimension-four Yukawa coupling and the dimension-six $(g-2)_\mu$ operators. Thus by closing the Higgs loop (and attaching photons for $(g-2)_\mu$) as in the left panel of Fig.~\ref{fig:quaddiv}, we can generate the dimension-four Yukawa coupling and dimension-six $(g-2)_\mu$ operators from quartic and quadratic divergences in this loop. (Note that the derivative $\slashed{D}^2$ in Eq.~(\ref{eq:LSout}) acts on the internal $H$).  Of course as always power-divergences are not calculable in the effective theory, but by the usual logic of naturalness, their presence is an indication for the size of the operator we can expect from dimensional analysis, and thus give an estimate for what we would get from the full UV theory.  

Indeed, we do obtain a dimension-four muon Yukawa operator
as  confirmed by the full UV theory computation resulting in Eq.~(\ref{eq:mass}). 
But the naturalness expectation for dimension-six $(g-2)_\mu$ operators is false;
the correction is exactly zero in the full UV theory.  The very low-energy effective field theorist thus sees that there is a dimension-eight operator breaking the relevant chiral symmetries, and also dimension-four Yukawa of about the right size expected from the quartically divergent estimate, but that the dimension-six $(g-2)_\mu$ operator is absent. Note that the cancellation happens from the contribution around $m_L$, which is far above the cutoff $m_S$ of the very low-energy effective theory. This is a concrete realization of the slogan that ``power divergences are absent for UV reasons, far above the naive cutoff of the effective theory", which is sometimes invoked to motivate how mysterious UV phenomena at the Planck scale might change the naturalness estimates for the Higgs mass or the cosmological constant at far lower scales. 

The surprise is perhaps more acute to the effective field theorist who lives at energy scales between $m_S$ and $m_L$. Integrating out $L$ at tree-level generates the operator
\begin{align}
\label{eq:Lout}
    {\cal L}_{\rm eff} = \frac{Y_R Y_V}{m_L^3} S H \slashed{D}^2 (H e^c) + \frac{Y_R Y_V'}{m_L^2} \overline{S} H \slashed{D} (H e^c).
\end{align}
Closing the Higgs and $S$ loop with  the Yukawa coupling $Y_L$, a dimension-six operator is generated around the energy scale $m_S$. This is a fully calculable IR contribution to $(g-2)_\mu$. But in this effective theory, there is also a UV matching contribution to $(g-2)_\mu$, from integrating out physics above the scale $m_L$, that cancels the calculable IR contribution exactly. To the effective field theorist living between $m_S,m_L$, this looks like a ``UV-IR connection/conspiracy", again of the sort sometimes hoped for in connection with the hierarchy and cosmological constant problems. But again everything has a very simple explanation, following from the ``total derivative phenomenon'' in the full theory. 

In our computation of the loop integral, we took the Higgs mass $m_H^2$ to be zero, as appropriate for leading effects in the effective theory far above the weak scale. A non-zero $m_H^2$ does give a non-vanishing integral. This is understood as the renormalization group equation (RGE) correction to the dimension-six $(g-2)_\mu$ operator from the product of a dimension-eight operator in Eq.~(\ref{eq:LSout}) and the dimension-two $m_H^2 |H|^2$ Higgs mass operator. One should, however, consistently include the correction from the Higgs quartic $\lambda|H|^4$ to a dimension-eight $(g-2)_\mu$ operator with additional $|H|^2$. Adding them up as in the middle panel of Fig.~\ref{fig:quaddiv}, the correction is proportional to $2 \lambda|H|^2 + m_H^2$ and vanishes at the minimum of the Higgs potential. This is not an accident, since at the minimum, the mass of the would-be charged Nambu-Goldstone scalar vanishes, and as we have seen, the one-loop contribution from a massless charged scalar vanishes. 

The massless scalar inside the loop and $f(0)\to 0$, which result in the vanishing $(g-2)_\mu$ at the leading order, are an important consequences of our assumption--motivated by anthropic considerations--that there are no scalars beyond the Higgs running in the loops for $(g-2)_\mu$.
If the scalar inside the loop in Fig.~\ref{fig:dim6} is a new scalar field $\eta$, the one-loop correction is non-zero. Such a setup is considered in~\cite{Freitas:2014pua,Baker:2021yli}.  The top two diagrams are generically dependent on different Yukawa couplings, and the partial cancellation in Eq.~(\ref{eq:integral2}), which leads to  $f(0)\to 0$, no longer occurs. This is reflected in the fact that an operator without derivative $(\ell \eta^\dag) \eta H e^c$ does not identically vanish.
In the bottom diagram, $f(0)$ is still zero, because the tree-level exchange of $S^c$ and $L$ does not generate the operator $(\ell \eta^\dag) \eta H e^c$.
Diagrammatically, this is due to the absence of the chirality flip in the $S^c$ and $L$ propagators. Perhaps a more interesting way to understand this is to observe that when $Y_V =0$ and the scalars are treated as non-dynamical fields, the theory has a symmetry under which $\e^c$ and $L^c$ are simultaneously shifted by a constant (also another where $(\ell, S)$ are shifted). Explicitly, putting the the scalars to their zero modes, the mass terms are written as $S^c(M_S S + Y_L \langle \eta \rangle \ell) + L(M_L L^c + Y_R \langle \eta \rangle e^c) + Y_V^\prime S^c \langle H \rangle L$, and so we have a shift symmetry $\ell \to \ell + \xi M_S, S \to S - \xi Y_L\langle \eta \rangle$, and a similar symmetry on $(e^c,L^c)$. Thus, while all the chiral symmetries are broken, this shift symmetry prohibits the generation of a mass term for $\ell, e^c$ when the heavy leptons are integrated out, guaranteeing that $f(0) = 0$. Although $f(0)=0$, since the scalar inside the loop is massive, non-zero $(g-2)_\mu$ is generated at the leading order, but is suppressed by $m_{\eta}^2/ m_{S,L}^2$ for $m_{\eta}^2 \ll m_{S,L}^2$.

\subsection{One-loop correction to dimension-eight operator}

\begin{figure}[t]
\begin{center}
 \includegraphics[width=0.48\textwidth]{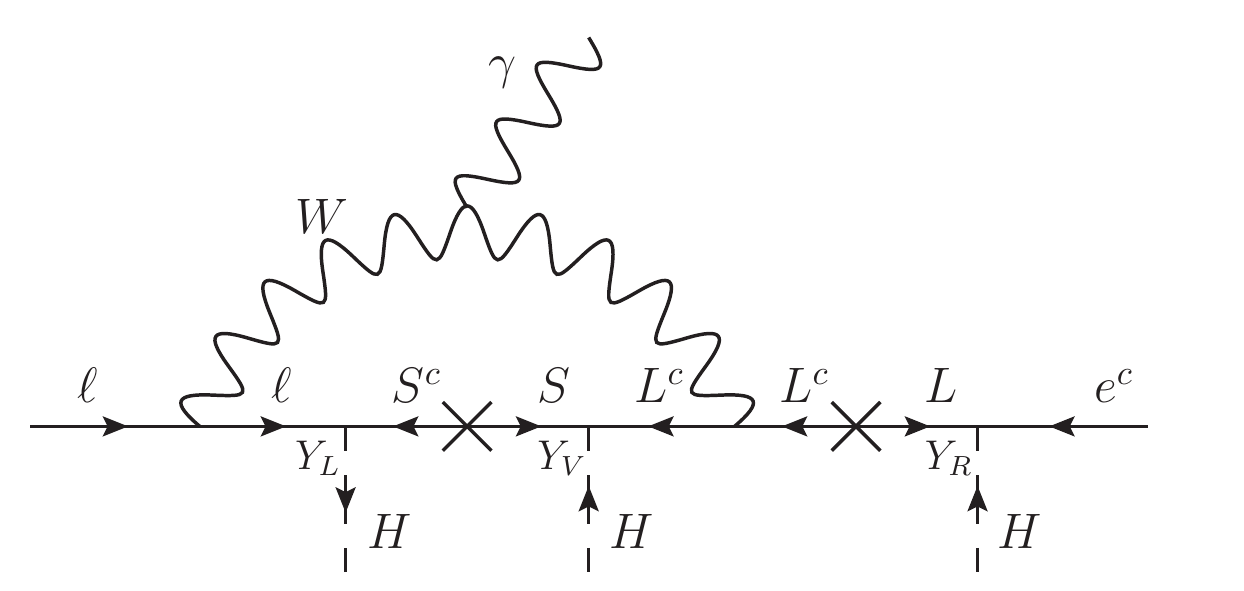}
 \includegraphics[width=0.48\textwidth]{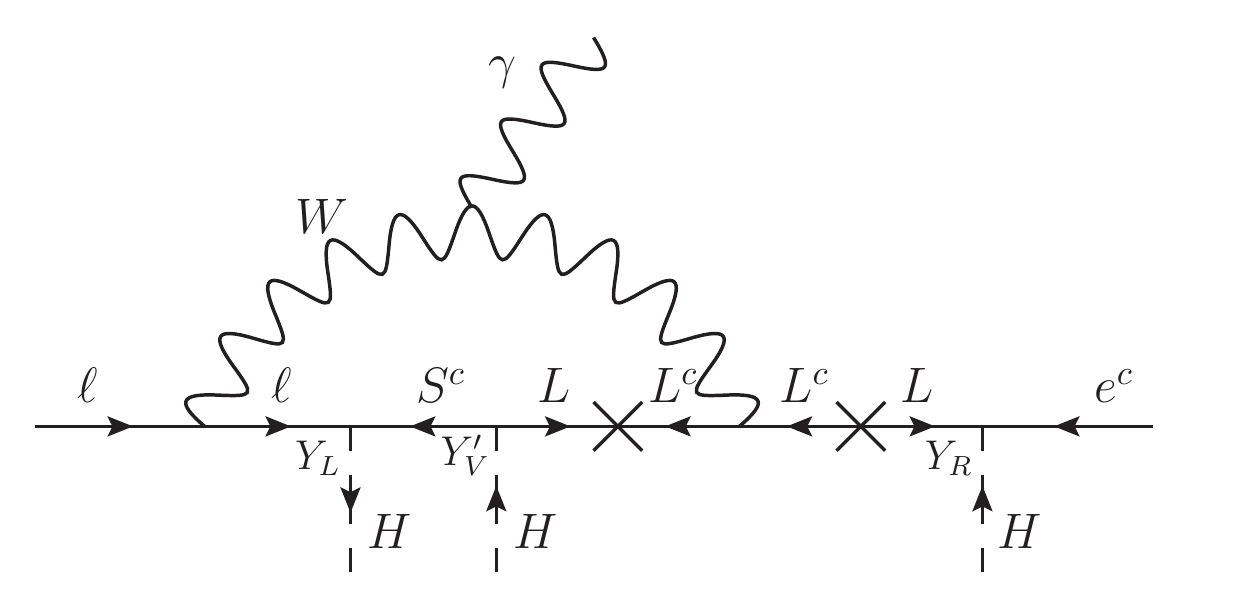}
 \caption{
 Diagrams that generate dimension-eight $(g-2)_\mu$ operators in the full UV theory.
}
\label{fig:dim8}
\end{center}
\end{figure}

We next consider dimension-eight operators $|H|^2 H \ell \sigma^{\mu\nu} e^c F_{\mu \nu}$ and $|H|^2 H \ell D^2 e^c$.
In the full UV theory, the operators are generated by the diagrams shown in Fig.~\ref{fig:dim8}.
In the low energy effective theory after integrating out $L$ and $S$, the correction is understood as the RGE correction from the dimension-eight operator in Eq.~(\ref{eq:LSout}) to the dimension-eight $(g-2)_\mu$ operators as is shown in the right panel of Fig.~\ref{fig:quaddiv}. For $m_S \ll m_L$, we obtain
\begin{align}
 \Delta a_\mu \simeq \frac{6 Y_LY_R}{16\pi^2} \frac{m_W^2 v m_\mu}{m_S^2 m_L^2} \left(Y_V \frac{m_S}{m_L} + Y_V' \right) {\rm log}\frac{m_S^2}{m_W^2}.
\end{align}
Note that there should not be a RGE correction at the energy scales between $m_L$ and $m_S$ and hence a log-factor ${\rm log}(m_L/m_S)$ is absent for the following reason. After integrating out $L$, we obtain dimension-seven and six operators in Eq.~(\ref{eq:Lout}). In concert with the marginal and relevant operators in the model, by dimensional analysis these operators can not generate dimension-eight operators under the RG. 

Since $m_S$ and/or $m_L$ are not much above $m_W$, we need to go beyond the leading-log approximation and compute the full one-loop contribution to $(g-2)_\mu$. We work in unitary gauge and compute the diagrams shown in Fig.~\ref{fig:dim8}. From our previous discussions in the Higgs picture above the weak scale, the contribution from the longitudinal component of the $W$ bosons with high momenta should vanish. In fact we find that this contribution vanishes identically at all loop momenta, and only the transverse component of the $W$ boson propagator contributes in unitary gauge. 
The final correction to $(g-2)_\mu$ is
\begin{align}
\label{eq:gm2}
    \Delta a_\mu =&\frac{6 Y_LY_R}{16\pi^2} \frac{m_W^2 v m_\mu}{m_S^2 m_L^2} \left(Y_V \frac{m_S}{m_L} + Y_V' \right)F\left(\frac{m_S^2}{m_W^2},\frac{m_L^2}{m_W^2}\right),\nonumber \\
    F(x,y) \equiv& \frac{x^3 y {\rm log}x}{(y-x)(x-1)^3} + \frac{x y^3 {\rm log}y}{(x-y)(y-1)^3 } - \frac{x y (3 xy -x-y-1)}{2 (y-1)^2(x-1)^2} > 0.
\end{align}

As can be seen from Eqs.~(\ref{eq:mass}) and (\ref{eq:gm2}), after choosing the signs of the parameters so that $\Delta a_\mu>0$, the correction to the muon mass is negative, $\Delta m_\mu<0$. This means that one cannot obtain the muon mass solely from the radiative correction at the energy scales below $\Lambda$ while explaining the deviation of $(g-2)_\mu$.
A non-zero muon Yukawa $\ell e^cH$ or a mass term $\ell L^c$ are required as the boundary values at the scale $\Lambda$, which may come from radiative corrections as we will discuss later.

\subsection{Two-loop correction to dimension-six operator}
Since the leading one-loop contribution to $(g-2)_\mu$ is suppressed by $m_W^2/m_{L,S}^2$, two-loop corrections without the suppression dominates over the one-loop correction for $m_{S,L} \gg m_W$. We will now examine the leading two-loop correction to the dimension-six operator--coming from the top Yukawa couplings--although as we will see, in the parameter range that can explain the observed deviation of $(g-2)_\mu$ this correction turns out to be numerically significantly smaller than the dimension-eight one-loop contribution.  

The top Yukawa correction arises from diagrams adding a top-loop to Fig.~\ref{fig:dim6}, with the photon attached to the left of the top loop, to the tops in the loop, and to the right of of the top loop. Since the electromagnetic current is not renormalized, the sum of these correction is simply proportional to the top-loop correction to the Higgs wavefunction renormalization. This amounts to modifying the integrand for $(g-2)_\mu$ in Eq.~(\ref{eq:int}) as
\begin{align}
    \int du f'(u) \left( 1 + \frac{3 y_t^2}{16\pi^2} {\rm log}u \right).
\end{align}
Because of the extra log-dependence, the integral no longer vanishes and gives a finite contribution,
\begin{align}
(\Delta a_\mu)_{\rm top} =& - \frac{3 y_t^2Y_RY_L}{(16\pi^2)^2} \frac{v m_\mu}{m_L^2 - m_S^2}  \left(Y_V \frac{m_S}{m_L} + Y_V' \right){\rm log}\left(\frac{m_L^2}{m_S^2}\right).
\end{align}

Although the correction involves a factor of ${\rm log}(m_L/m_S)$, this is {\it not} from the RGE correction between $m_L$ and $m_S$. This is evident from the form of the integrand that is shown in Fig.~\ref{fig:integrand}; the integrand is peaked at $m_S^2$ and $m_L^2$ with opposite signs. The contributions from these two peaks exactly cancel without the extra log-dependence. With the extra log-factor, the cancellation is imperfect and results in a factor ${\rm log}m_L-{\rm log}m_S = {\rm log}(m_L/m_S)$. The non-zero correction proportional to ${\rm log}(m_L/m_S)$ should be thus understood as the sum of two threshold corrections (whose magnitudes logarithmically depend on the energy scale) at the scales $m_L$ and $m_S$ rather than as the RGE effect between $m_L$ and $m_S$.

The correction does not involve ${\rm log} (m_{L,S}/m_W)$, in contrast to the ${\rm log}(m_{L,S}/m_W)$ enhancement in the dimension-eight $(g-2)_\mu$ operator, for the following reason.
For such a factor to arise, the dimension-six $(g-2)_\mu$ operators must be generated from other dimension-six operators through the RGE after integrating out $L$ and $S$. However, the operator $\ell \slashed{D}^2 e^cH$ can be removed by a field redefinition shifting of $\ell$ proportional to $\slashed{D}(\overline{e^c} H^\dag)$, and does not contribute to the RGE. After the shift, a dimension-six operator proportional to the muon Yukawa coupling remains, but $(g-2)_\mu$ from that coupling is proportional to $m_\mu^2$ and is negligible.

Two-loop corrections from other interactions, such as the electroweak gauge interactions and the Higgs quartic coupling, also generate $(g-2)_\mu$. For the same reason as the top loop, a factor ${\rm log} (m_{L,S}/m_W)$ is absent. We find that these corrections are negligible in the parameter region that can explain the deviation of $(g-2)_\mu$ unless a numerical factor in addition to the loop factor $1/(16\pi^2)^2$ and a log-enhancement ${\rm log}(m_L/m_S)$ is more than 30, and neglect them. It will be, however, of interest to perform a full two-loop computation.

All of this discussion highlights the importance of having a free Higgs propagator for the cancellation we have found. Indeed, we could imagine general theory where the Higgs coupling is replaced with a general operator in a CFT, with some anomalous dimension $\gamma_h$ so that the propagator is replaced by $1/k^{2 + \gamma_h}$. The integral no longer vanishes, and we would instead get a contribution proportional to $\left[\left(M_L/M_S\right)^{\gamma_h} - 1\right]$. This is yet another manifestation that our cancellation is not a scale-by-scale phenomenon, and depends on details such as the fact that the Higgs is close to being free to scales above that of the heaviest of the new lepton masses. 

\section{Phenomenological implications}

\begin{figure}[!t]
\begin{center}
 \includegraphics[width=0.49\textwidth]{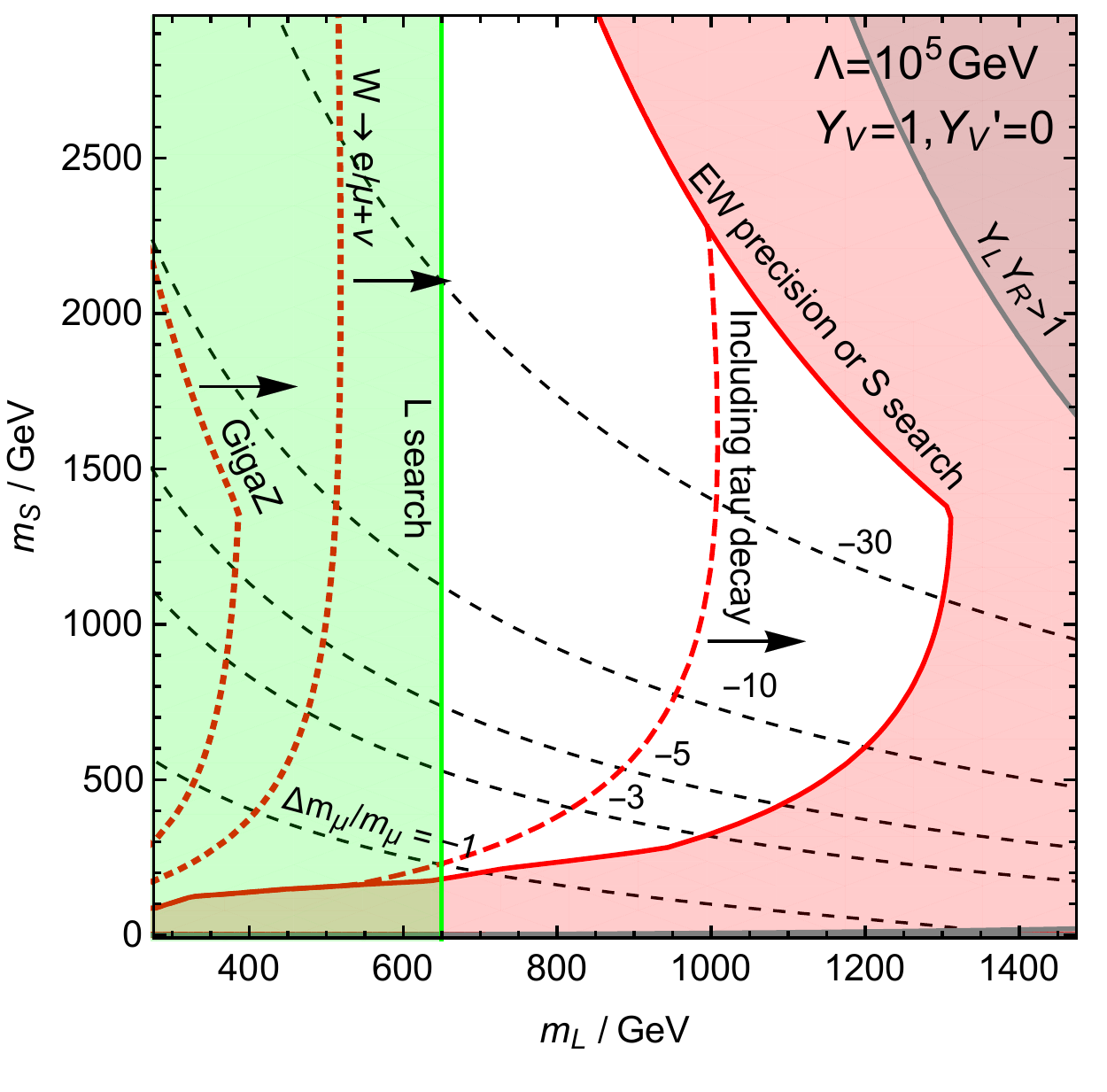}
 \includegraphics[width=0.49\textwidth]{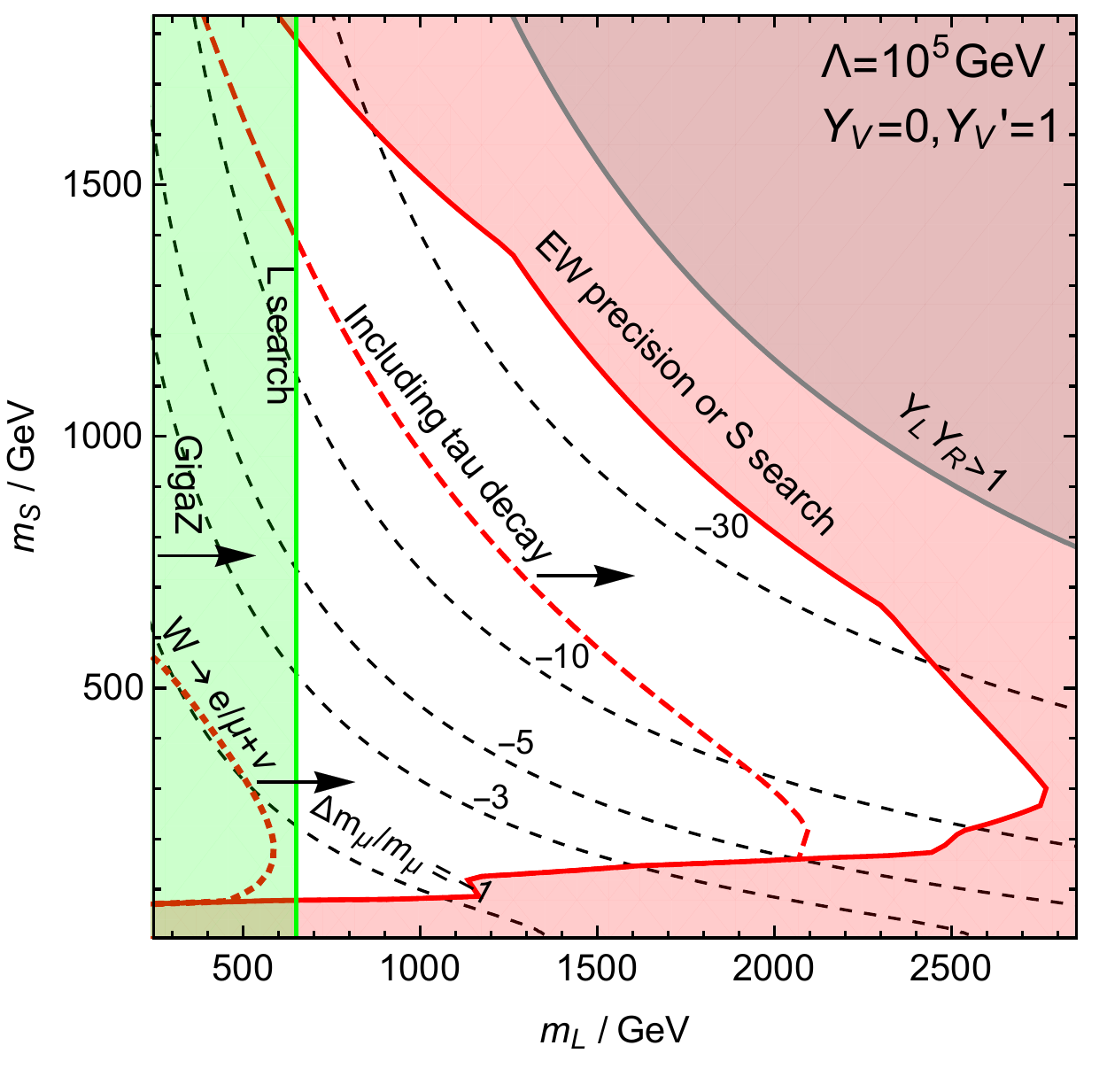}
 \caption{
 Constraints on the model for a fixed $\Delta a_\mu = (2.51-0.59)\times 10^{-9}$. The part of parameter space without significant tuning for obtaining the muon mass can be probed by the LHC search for the doublet vector-like lepton $L$. {\it All} of the parameter space can be probed by future lepton colliders.
}
\vspace{2.5cm}
\label{fig:gm2}
 \includegraphics[width=0.49\textwidth]{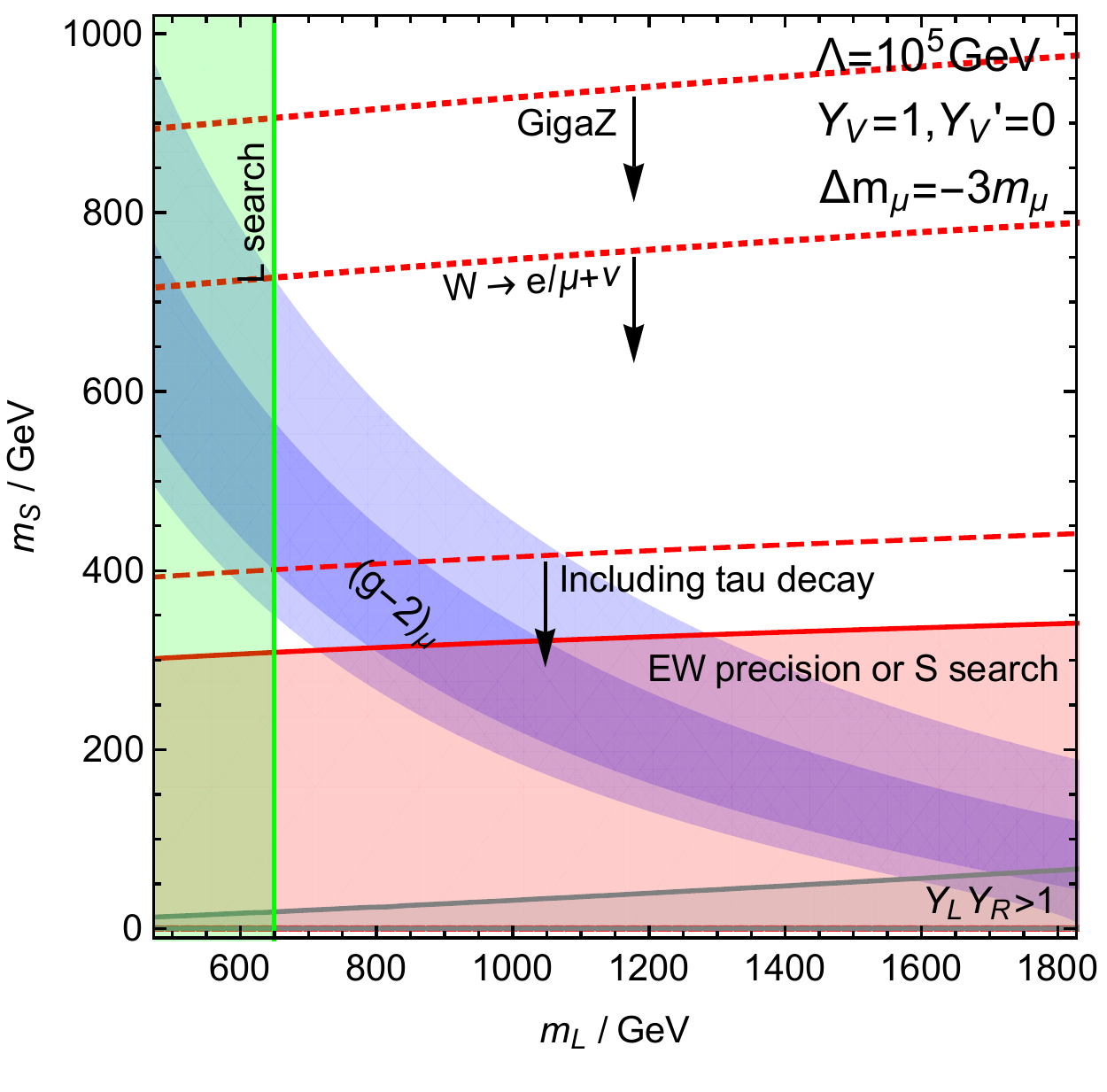}
 \includegraphics[width=0.49\textwidth]{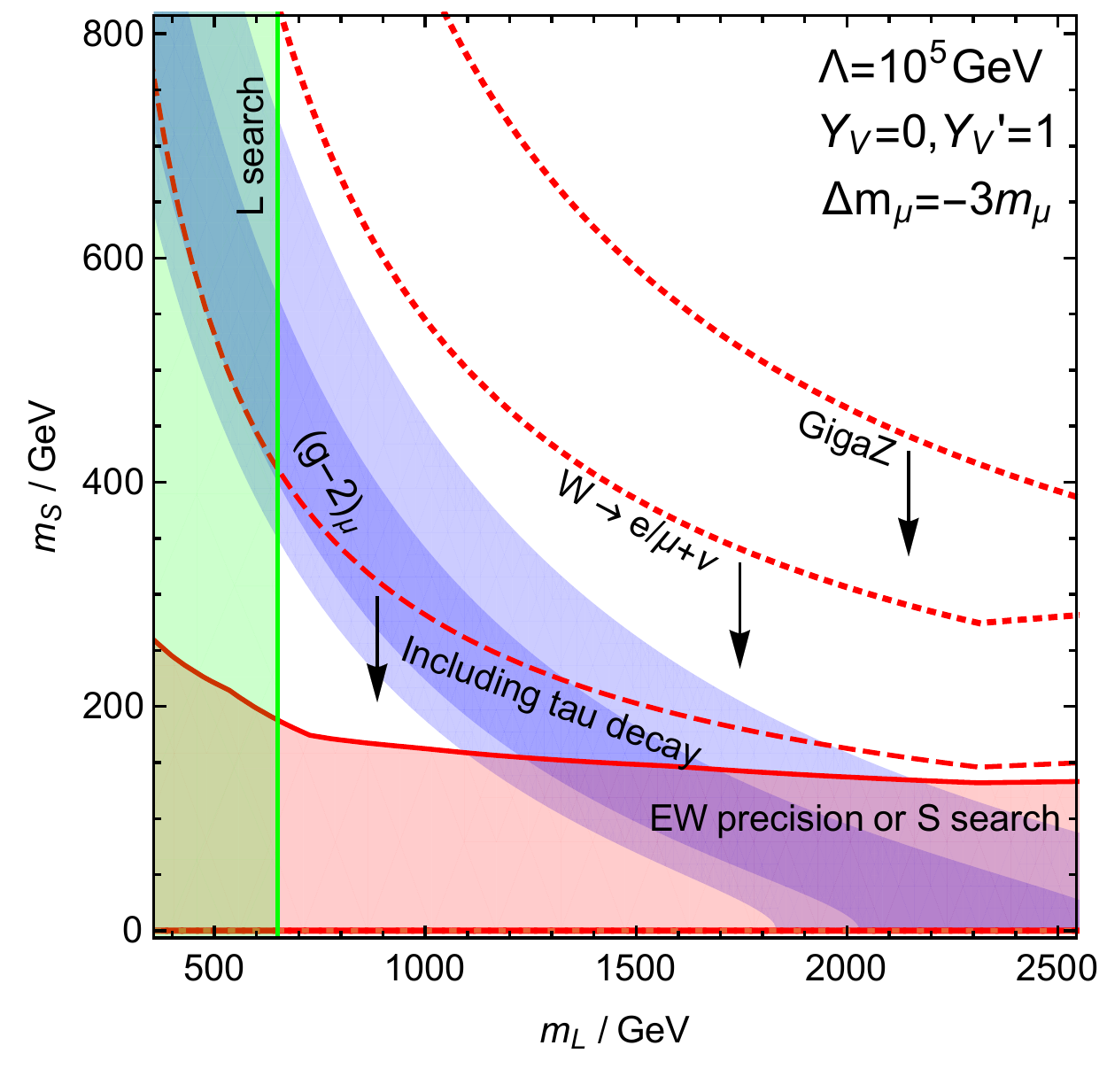}
 \caption{
 Constraints on the model for a fixed $\Delta m_\mu = - 3 m_\mu$.
}
\label{fig:mass}
\end{center}
\end{figure}

In Fig.~\ref{fig:gm2}, we show the constraints on $m_L$ and $m_S$, requiring $\Delta a_\mu = (2.51-0.59)\times 10^{-9}$.  We take $Y_V=1$ and $Y_V'=0$ in the left panel and $Y_V=0$ and $Y_V'=1$ in the right panel. In the gray-shaded region, $Y_L Y_R >1$, for which the Higgs potential becomes unstable below 10 TeV. The dashed black lines show the contours of $\Delta m_\mu/m_\mu$ assuming $\Lambda = 10^5$ GeV; lowering $\Lambda$ reduces $\Delta m_\mu/m_\mu$.

The doublets $L$ and $L^c$ are produced at colliders and decay into the SM leptons and $W$, $Z$, or Higgs bosons. The search for such signals in~\cite{Aad:2020fzq} assumes an $SU(2)_L$ triplet fermion, but we expect that the constraint for triplets is similar to that for doublets up to the three times larger cross section. We then obtain a bound $m_L > 650$ GeV, which is shown as the green-shaded region in Fig.~\ref{fig:gm2}.
Note that this search assumes direct decay of the triplet into SM particles, while in our set-up, $L$ can first decay into $S$ and $H$, and $S$ can decay into $\ell$ and $H$, but we expect the constraints to be similar. The sensitivity can be improved by utilizing extra leptons or looking for peaks at the invariant mass of the intermediate $S$.
High-Luminosity LHC can probe vector-like leptons with a mass $m_L < 1250$ GeV even if they dominantly couple with the third generation leptons~\cite{Bhattiprolu:2019vdu}. Because of the dominant coupling with the muon, we expect better sensitivity in our set-up; the parameter space without significant tuning in $m_\mu$ will thus be incisively probed by future searches at the LHC. 

The parameter space of the model can be also probed by the precise measurements of $Z$ and $W$ boson decays.
After integrating out the vector-like leptons, we obtain effective operators
\begin{align}
 \frac{Y_L^2}{m_S^2} (\overline{\ell} H) i \bar{\sigma} D (\ell H^\dag) +  \frac{Y_R^2}{m_L^2} (\overline{e^c} H) i \bar{\sigma} D (e^c H^\dag).
\end{align}
These effective operators modify the coupling of $Z$ and $W$ bosons with $\mu$ and $\nu_\mu$,
\begin{align}
{\cal L} \supset 
&\left(\frac{g}{\sqrt{2}}W_\rho^+   \bar{\nu}_\mu\gamma^\rho \left(1 + \delta g_L^W\right) P_L  \mu + {\rm h.c.}\right)  \\
&+ \frac{g}{c_W}Z_\rho  \left( \bar{\mu}\gamma^\rho \left(s_W^2 + \delta g_R^{Z\mu}\right) P_R  \mu +  \bar{\nu}_\mu \gamma^\rho \left( \frac{1}{2} + \delta g_L^{Z\nu}\right) P_L \nu_\mu  \right), \nonumber  \\
 &\delta g_L^W = \delta g_L^{Z\nu} =- \frac{Y_L^2 v^2}{2m_S^2},~
\delta g_R^{Z\mu}= -\frac{Y_R^2 v^2}{2m_L^2}. \nonumber
\end{align}
The $W$-$\mu \nu$ and $Z$-$\nu$ couplings decrease since the SM-like $\nu_\mu$ contains a small fraction of a singlet $S$ that does not couple to the gauge bosons.
The $Z$-$\mu$ coupling also decreases since the SM-like right-handed $\mu$ contains a small fraction of a doublet $L^c$ whose coupling to the $Z$ boson is opposite to that of a singlet $e^c$.
The present constraint from the global electroweak fit is~\cite{Breso-Pla:2021qoe}
\begin{align}
    \frac{Y_L^2 v^2}{m_S^2} < 1.6 \times 10^{-2},~~\frac{Y_R^2 v^2}{m_L^2} < 5.6\times 10^{-3}.
\end{align}
Direct collider search for $S$ produced by the mixing of it with the SM neutrinos also puts an upper bound on $Y_L v/m_S$~\cite{Sirunyan:2018mtv}.
These indirect and direct bounds on $Y_L$ and $Y_R$ as well as the requirement of $Y_{L,R}<1$ exclude the red-shaded region in Fig.~\ref{fig:gm2}. 

The modified $W$ boson coupling also affects tau decay. In our model, $\Gamma(\tau \rightarrow \mu \nu \bar{\nu})/\Gamma(\tau \rightarrow e \nu \bar{\nu})$ is smaller than the SM prediction of $0.9726$, while the measured value of it, $0.9762 \pm 0.0028$, is above the SM prediction~\cite{Zyla:2020zbs}. Assuming that there is no other new physics that can affect the lepton universality of tau decay, this puts a strong upper bound on $Y_L v/ m_S$. We require that our model does not give worse fit than the SM by more than $2 \sigma$, assuming that the error of the measurement can be approximated by a Gaussian distribution even outside the quoted error bar. We then obtain
\begin{align}
    \frac{Y_L^2 v^2}{m_S^2} < 5.7 \times 10^{-3},
\end{align}
which is shown by the red-dashed line without shading in Fig.~\ref{fig:gm2}.
If we instead require that $\Gamma(\tau \rightarrow \mu \nu \bar{\nu})/\Gamma(e \rightarrow \mu \nu \bar{\nu})$ in our model falls in the face value of $0.9762 \pm 0.0028$, nearly all of parameter space is disfavored except for that with small $m_L$ or $m_S$, because the model predicts a deviation from the SM opposite to what was observed.
Note that this constraint arises if we imagine that only the coupling to muons is modified; we will return to discuss this point in a moment when we introduce a larger setting for these models, motivated by the radiative generation of fermion masses, where the strong tau decay constraint can be easily eliminated.

Future lepton colliders can probe the full parameter space of the theory. Lepton universality in $W$ decays can be measured with an accuracy of $6\times 10^{-4}$ by $e_L^-e_R^+\rightarrow W^-W^+$ with $\sqrt{s}=250$ GeV and the integrated luminosity of $0.9$ ab$^{-1}$ at the International Linear Collider~\cite{Fujii:2019zll}. The branching ratio of $W$ into $\mu\nu$ can be measured with a similar accuracy. A GigaZ factory can measure the $Z\mathchar`-\mu$ coupling with an accuracy of $2\times 10^{-4}$~\cite{Fujii:2019zll}. Similar sensitivities for precision measurements of $W/Z$ couplings are expected at the Circular Electron Positron Collider~\cite{CEPCStudyGroup:2018ghi,CEPCEW}. These measurements cover {\it all} of the viable parameter regions for explaining $(g-2)_\mu$, as indicated by the red dashed lines with labels ``$W\rightarrow e/\mu+\nu$" and ``GigaZ" in Fig.~\ref{fig:gm2}.

In Fig.~\ref{fig:mass}, we instead fix the Yukawa couplings so that $\Delta m_\mu = -3 m_\mu$. Inside the blue band, $\Delta a_\mu = (2.51 \pm 0.59)\times 10^{-9}$. Other constraints and prospects are the same as those in Fig.~\ref{fig:gm2}. Again, future lepton colliders can probe the entire viable parameter region.

\section{Summary and Discussion}
\label{sec:discussion}
We studied a simple model that can explain the observed deviation of the muon anomalous magnetic moment from the (data-driven) SM prediction. The model only introduces new vector-like fermions and does not suffer from naturalness problems beyond that of the SM Higgs, compatible with an anthropic explanation for the small weak scale.

Because of the absence of new scalars around the TeV scale, the model can be easily embedded into a framework with a little hierarchy. Perhaps the best-motivated are supersymmetric theories with squarks and sleptons around the $100-1000$ TeV scale~\cite{Giudice:1998xp,ArkaniHamed:2004fb,Giudice:2004tc,Wells:2004di,Hall:2011jd,Ibe:2011aa,Arvanitaki:2012ps,ArkaniHamed:2012gw}.
The large scalar mass is not only consistent with the observed Higgs mass of 125 GeV, but also is free from the flavor, gravitino, and moduli problems. Because of the large scalar mass, correction to $(g-2)_\mu$ are dominated by loops of the vector-like fermions and the $W$ boson we have studied. The cut-off scale $\Lambda$ is identified with the mass of the heavy scalars. 

The model with $Y_V'=0$ can be embedded into a theory where the muon mass (as well the tau and electron masses) can be generated radiatively. We may introduce a $U(1)$ symmetry with charges $\ell(-2)$, $e^c(0)$, $L(0)$, $L^c(1)$, $S(-1)$, and $S^c(2)$. The symmetry forbids a tree-level muon Yukawa coupling $\ell e^cH$ and $Y_{V}'$. We introduce a soft breaking of the symmetry by $m_L$ and $m_S$. The same spurion with an opposite $U(1)$ charge can give a mass term $\ell L^c$, but that may be forbidden by the holomorphy in supersymmetric theories. The muon Yukawa coupling is given by the radiative correction below the scale $\Lambda$ shown in Eq.~(\ref{eq:mass}) and a threshold correction at $\Lambda$, namely, the soft scalar mass scale. The latter must dominate over the former and flip the sign of the muon mass to explain the sign of the deviation of $(g-2)_\mu$, as can be seen from Eqs.~(\ref{eq:mass}) and (\ref{eq:gm2}). Note that this gives an extra motivation, beyond the simple story for supersymmetry breaking and successful prediction for the Higgs mass, for only a ``mini-split" spectrum, to avoid too large logarithmic enhancement of the running contribution to the muon Yukawa.
The needed threshold correction can arise,  for example, by a Higgsino-scalar $SS^c$ loop with a moderately large $Bm_S$ term.

The electron and tau masses can be also radiatively generated by introducing extra vector-like leptons that have the same charge as $LL^c$ and $SS^c$ and couple to electrons and taus in the same manner as in Eq.~(\ref{eq:L}).
Because of the dependence of the radiatively generated charged lepton masses on the cubic power of the Yukawa couplings in Eq.~(\ref{eq:L}), $O(0.1\mathchar`-1)$ hierarchy in the Yukawa couplings between generations can explain the charged lepton mass hierarchy.
It is also possible to extend the model so that the electron mass arises at two-loop level by generating a coupling or a mass in Eq.~(\ref{eq:L}) by one-loop radiative corrections.

Sadly, a lepton flavor symmetry should be introduced to suppress flavor changing decays such as $\mu \rightarrow e \gamma$. The neutrino mixing, which violates the lepton flavor symmetry, may arise from soft breaking of the symmetry. For example, we may consider a see-saw mechanism~\cite{Minkowski:1977sc,Yanagida:1979as,Mohapatra:1979ia,GellMann:1980vs} with the lepton symmetry and the lepton flavor symmetry softly broken by right-handed neutrino masses. We may also introduce soft breaking by the Majorana mass of $S$ and/or $S^c$. These lepton flavor violations break the lepton number by two units and do not generate a lepton flavor violation in the charged lepton sector.

If the vector-like leptons for the electron Yukawa is as light as those for the muon Yukawa, to avoid too large an electron electric dipole moment (EDM), an (approximate) CP symmetry must be introduced, to ensure phase alignment between the electron mass and dipole moment operators at the $10^{-4}$ level.
The electron EDM can be suppressed without the CP symmetry if the masses of the vector-like fermions for the electron Yukawa are heavier. In this limit the two-loop correction involving the top Yukawa coupling dominates the EDM, and the required masses of the vector-like fermions are $O(100)$ TeV. Note that in the model with $Y^\prime_V = 0$, the induced Yukawas are supressed by $m_S/m_L$, thus keeping the electron-singlet $S$ light but making the electron-doublet $L$ heavy, near $\sim 100$ TeV, would generate the $e\mathchar`-\mu$ Yukawa hierarchy while evading a large EDM, even with no suppression of the CP phase. 

It is clearly more attractive to introduce vector-like leptons for all three generations than simply one for the muon, and in concert with supersymmetry at the $\sim 100$ TeV scale, this lets us build an interesting model for the radiative origin of the lepton masses. The extra vector-like fermions also allow us to easily evade the strong constraints on the model from deviations in lepton universality from $\tau$ decay we alluded to above. This is because we can also expect similar-size deviations in the couplings of the $W$ to electrons and muons (these are controlled by the left-handed $Y_L$ Yukawa couplings and the $M_L$ masses, that can be comparable, while the right-handed couplings $Y_R$ could be smaller for electrons as part of the explanation of the $m_e/m_u$ hierarchy). Thus the shift in the branching ratios between muons and electrons in $\tau$ decay can be altered to have either sign, eliminating the strong constraint on the parameter space of the model with only vector-like lepton partners of the muon. Even for this case, the current constraint on the $W\mathchar`-\mu \nu$ coupling from the global electroweak fit in~\cite{Breso-Pla:2021qoe} remains similar, so a GigaZ factory can still fully probe the parameter space of the model.

In order to preserve the success of supersymmetric gauge-coupling unification, the new vector-like leptons should  be embedded into complete ${\bf {5}} + {\bf \bar{5}}+{\bf 1}+{\bf 1}$ multiplets in $SU(5)$, by supplementing the model with $D$ and $D^c$ with the  opposite and same gauge charges as the right-handed down quark. Unlike models with hyper-charged $SU(2)_L$ singlets, we do not need ${\bf 10}$ in $SU(5)$, and so the perturbativity of the gauge couplings can be easily maintained up to the GUT scale.
If the $U(1)$ symmetry discussed above acts on the full ${\bf 5}$-plets, a tree-level strange Yukawa coupling is forbidden. Also, because of the absence of the colored Higgs, quantum corrections to the strange mass are absent at the TeV scale. But the mini-split supersymmetry spectrum offers other simple sources of radiative Yukawas, from integrating out squarks with flavor-violating soft masses, or via the addition of additional vector-like matter at the $100-1000$ TeV scale. Alternately the $U(1)$
symmetry can be taken to act only on the leptons, in which case a tree-level Yukawa for the down-type quarks is allowed. 

Given the new colored states $D,D^c$ needed to preserve unification, it is tempting to try and explain the observed lepton non-universality of the $B$ meson decay~\cite{Aaij:2021vac} by the quantum correction from the box diagram involving the loop of $S$, $D$ and $H$. But this is easily seen to be impossible. In order to avoid the constraint from the $B_s\mathchar`-\overline{B_s}$ mixing~\cite{Amhis:2016xyh}, the Yukawa coupling $q D^c H$ must be small, forcing non-perturbatively large coupling to leptons in order to get a large enough correction to $B_s \to \mu^+ \mu^-$. To explain the $B$ anomaly, further extension of the model is required, see e.g.,~\cite{Kawamura:2019hxp}.

As we discussed in this note, in our simple model, the quantum correction to dimension-six $(g-2)_\mu$ operators vanishes at one-loop level, lowering the required new physics mass scale. The viable parameter space can be fully probed by the LHC and planned future lepton colliders. Let us conclude by making some further simple observations on this interesting counterexample to Wilsonian naturalness we have encountered here.

Recall that the left-right dimension-six operators contributing to $(g-2)$ are of the form $\ell H {\cal O} e^c$ where ${\cal O}$ is either ${\cal O}=D^2$ or $\sigma^{\mu \nu}F_{\mu \nu}$. (Of course these operators are equivalent when $\ell$ and $e^c$ are on-shell, since $\slashed{D}^2 = D^2 + \sigma^{\mu\nu} F_{\mu \nu}$ and  $\slashed{D}^2 e^c = m^2 e^c$ on-shell). 
For simplicity, let us focus on a limit where the coupling $Y_V \to 0$, so only the bottom diagram in Fig.~\ref{fig:dim6} contributes to $(g-2)_\mu$. (A slight elaboration of the following discussion applies to the general case where both $Y_V,Y_V'$ are non-zero).  Note that with the Higgs set to its vacuum expectation value, the exchanged fermion is neutral, so the only photon insertion comes on the charged scalar line in the diagram, and thus it is obvious that the spinor contractions between $\ell_{\alpha}$ and $e^c_{\beta}$ is proportional to $\epsilon^{\alpha \beta}$. Therefore the dimension-six operator contributing to $(g-2)$ must be  $\ell H D^2 e^c$. Amusingly, this allows to compute $(g-2)$ using diagrams without any photons attached, instead just computing the coefficient of $p^2$ in the expansion of the one-loop correction to the left-right two-point function. 
This argument extends to corrections at all loop order for corrections that are represented as  blobs modifying the Higgs propagator or the Higgs-Higgs-photon vertex, since all such corrections have the same trivial spinor contraction structure. (This includes the case of the two-loop top Yukawa contribution we discussed above, where the correction involved only the Higgs wavefunction renormalization factor, just as expected from examining the two-point function without a photon attached). 

The argument also extends to the case where any number of photons are attached to the charged Higgs line; the only operators that can be generated are of the form $\ell H (D^2)^n e^c$, whose coefficients can be computed from diagrams with no photons attached, by looking at the coefficient of $(p^2)^n$ in the expansion of the left-right two-point function $A(p^2)$.
The surprise that $(g-2)$ vanishes at this order extends to an interesting statement about {\it all} operators $\ell H (D^2)^n e^c$, arising from the $(p^2)$-expansion of the two-point function. 

In fact the point is more general than the vanishing of the coefficient of the $(g-2)/D^2$ operators. To illustrate it in a general setting, consider a theory with two fermions $e, e^c$, a charged ``higgs" $h$, $N$ neutral fermions $\Psi_I$, and any number of other scalars $\sigma_a$. We have the yukawa couplings $(e h^\dagger) (Y^I \Psi_I) + (e^c h)(Y^{c I} \Psi_I)$. We also have mass term $M^{IJ} \Psi_I \Psi_J$, Yukawas $\kappa^{I J a} \Psi_I \Psi_J \sigma_a$, and any masses and self-interactions for the $\sigma_a$. Let us now look at the effective action to leading order in the couplings $Y,Y^c$, but to all orders in the other couplings. We have terms like $(e h^\dagger Y) [{\bf B} + {\bf C} D^2 + {\bf D} (D^2)^2 + ....] (Y^c e^c h)$, as well as $\frac{1}{8 \pi^2}(e Y)[{\bf \tilde{A}} + {\bf \tilde{B}} D^2 + {\bf \tilde{C}} (D^2)^2 + ...] (Y^c e^c)$. 
Here $\bf{B},{\bf C},{\bf D},...$ and ${\bf \tilde{A}}, {\bf \tilde{B}}, {\bf \tilde{C}},\dots$ are $N$-by-$N$ $U(N)$ charged matrices, where we use the same letters to denote matrices of the same mass dimensions.

Now all these matrices are built out of the $U(N)$ charged matrices $M^{IJ}$, $\kappa^{IJa}$ etc.; the $\kappa$'s are dimensionless, so these matrices can be all mixed up. So a priori, while the matrices ${\bf B},{\bf C},{\bf D},...$ and ${\bf \tilde{B}}, {\bf \tilde{C}}, {\bf \tilde{D}}$ have the same mass dimension, there is no reason for them to be related just on symmetry grounds.  

This is where the ``total derivative phenomenon" yields the ``surprise". In fact up to numerical factors the matrix ${\bf \tilde{B}} = {\bf B}$, ${\bf \tilde{C}} = {\bf C}$, etc. We can see this because $[{\bf B} + {\bf C} D^2 + {\bf D} (D^2)^2 + ...]$ is really the expansion of some matrix ${\bf f}(k^2)$--the two-point function of the $\Psi$ fermions--around $k^2 = 0$, so that ${\bf B} = {\bf f}(0), {\bf C}={\bf f}'(0)$, etc.  But then the second set of operators arises simply by performing the loop integral
\begin{eqnarray}
8 \pi^2 {\bf A}(p^2)  &=& \int \frac{d^4 k}{k^2} {\bf f}((k+p)^2) \nonumber \\ &\propto&\int du {\bf f} \nonumber \\ &&+ p^2 \int du \left( {\bf f}^\prime + \frac{1}{2} u {\bf f}^{\prime \prime} \right) \nonumber \\ & &+  (p^2)^2 \int du \left(\frac{1}{2} {\bf f}^{\prime \prime} + \frac{1}{2} u {\bf f}^{\prime \prime \prime} + \frac{1}{12} u^2 {\bf f}^{\prime \prime \prime \prime} \right) \nonumber \\ &&+ \cdots .
\end{eqnarray}
Note that the integral for the coefficient of $p^2$, which we identify with ${\bf \tilde{B}}$, is a total derivative $\int du \frac{1}{2} (f + u f^\prime)^\prime =-\frac{1}{2}{\bf f}(0) = -\frac{1}{2} {\bf B}$. Similarly the coefficient of $(p^2)^2$, which we identify with ${\bf \tilde{C}}$, is $-\frac{1}{6} \bf{f}^\prime(0) = -\frac{1}{6} {\bf C}$ and so on.  A similar statement holds in any even number of spacetime dimensions $D$. The integrand above is multiplied by an extra Jacobian factor of $u^{D/2 - 2}$, and thus all ``primed" matrices in the coefficients $(p^2)^m$ with $m \geq (D/2 - 1)$ involve integrals of the form $\int du u^r f^{(s)}$ with $s>r$, and hence integrating by parts up to a numerical factor equal the ``unprimed" matrices ocurring in the $(p^2)^{(m-D/2+1)} h^\dagger h$ operators. This does not happen for odd spacetime dimensions, where the Jacobian involves odd powers of $\sqrt{u}$ and the integration-by-parts argument fails. Note that in this discussion, we have assumed that $m_h^2 \to 0$. Expanding in powers of $m_h^2$, the leading correction to the coefficient of $p^2$ in four dimensions is $m_h^2 \int \frac{du}{u} {\bf f}^\prime$, which is not a total derivative; so the precise equality between the coefficients is broken at order $(m_h^2/M^2)$. Amusingly in higher dimensions, the total derivative structure allows us determine some of terms in the $m_h^2$ expansion. For instance in $D=6$, while the leading $p^2$ operator $\int du u {\bf f}^\prime$ is not a total derivative, the $m_h^2$ correction $m_h^2 \int du {\bf f}^\prime$ {\it is} a total derivative. And we stress again that all of this depends crucially on the higgs propagator being free; if the higgs was replaced with e.g. an operator with some anomalous dimension $\gamma_h$ in a CFT, there would be corrections to all these relations proportional the $\gamma_h$. 

In our $(g-2)$ example, the statement is that, even including arbitrarily many extra loops as blobs just on the fermion line of the diagram, the coefficients in the expansion of $(\ell H^\dagger) H (\slashed {D}^2)^{n-1} (H e^c)$ are equal (up to the same numerical factors) to those of the same mass dimension in the expansion of $(\ell H) (D^2)^n e^c$.  
 
There is then a separate, more familiar ``surprise" in our $(g-2)$ example, which we have already remarked in our discussion emphasizing the importance of $f(u \to 0) = 0$. Upon integrating out the heavy fermions at tree-level, while there are no symmetries to forbid the generation of the dimension-six operator $\ell H |H|^2 e^c$ that would contribute to the muon mass, this operator is actually not generated. The mechanical reason for this is trivial--putting the Higgs to its vacuum expectation value, we only have mixing with the neutral components of the heavy fermions, and so obviously no muon mass term can be generated. As discussed earlier, more formally, in whatever operator we generate the $SU(2)$ indices of $\ell$ are contracted with $H^\dagger$, so the only possible dimension-six operator would be $(\ell H^\dagger) (HH) e^c$, which vanishes due to the antisymmetric contraction of the $SU(2)$ indices on the Higgs. As we also discussed for the case of the $Y_V^\prime$ coupling, even if the Higgs appearing in the Yukawa coupling to the heavy fermions is different from the one coupling to the SM leptons--so that  dimension-six operator can be written down--we still do not generate it. This is because turning on the higgses only as background fields, there is a symmetry simultaneously shifting $e^c, L^c$ (and another shifting $\ell,S$). All of this this obviously also holds when any number of loops are added as blobs only to the heavy fermion line. Since this also sets the coefficient  of the $D^2$ operator, $(g-2)$ vanishes as well. 

In the language of our general example,  this shows us that there may be a simple reasons why $(Y {\bf B} Y^c) = 0$. When this happens, the coefficient of the $D^2$ operator (and hence of $(g-2)$) also vanishes. But the more interesting  ``surprise" of the total derivative phenomenon is that there is  precise relationship between the coefficients of two different sets of operators, when working at leading order in the couplings to the Higgs.    

It is also illuminating to return to one-loop using the language of our general example; here we ignore all dependence on the extra couplings of the fermions $\Psi_I$ to other fields, so the only couplings charged under the $U(N)$ symmetry are $Y^I,Y^{c I}$ and $M^{IJ}$, and we work to linear order in each of $Y,Y^c$. Consider for instance the fermion mass term $(e e^c)$; the $U(N)$ invariants of the correct mass dimension are $\left(Y M^\dagger \left( M M^\dag / {\rm Tr}\left[ M M^\dag \right] \right)^nY^c\right)$ etc.  Now look at dimension-five operators $(e h^\dagger h e^c)$; by dimensional analysis and $U(N)$ invariance, the matrix sandwiched between $Y,Y^c$ can be
$M^{-1}\left( M M^\dag / {\rm Tr}\left[ M M^\dag \right] \right)^n$ etc. Of course it is obvious just from looking at the fermion propagator that we get $M^\dagger (M M^\dagger)^{-1} = M^{-1}$. (The absence of
other operators
is equivalent to the mild tree-level ``surprise" that the muon mass operator was not generated in our model). Next consider the $(e^c D^2 e)$ operator generated at one-loop, again just by dimensional analysis and $U(N)$ invariance we can write down $M^{-1}$, $\frac{M^\dagger}{{\rm Tr}(M^\dagger M)}$, \dots. Anson Hook has given a very simple argument for why we must again get $M^{-1}$ at one-loop. If we do the computation rotating to the mass eigenstate basis, we get a finite contribution from each eigenstate of mass $m_i$, which by dimensional analysis is $1/m_i$. When rotated back to a general basis, this gives the operator $M^{-1}$. The same argument works for the $(e h^\dagger h e^c)$ operator induced at tree-level. So we see that the matrices ${\bf B}, {\bf \tilde{B}}$ are both proportional to $M^{-1}$. This gives another explanation of why ${\bf B}, {\bf \tilde{B}}$ are proportional to each other. Note this is a one-loop statement: the simplicity of the dependence of the one-loop result on the masses $m_i$ is crucial to this argument; if we had instead $1/m_i g(m_i^2)$ with some non-trivial functions $g$, this would give $M^{-1} g(M M^\dagger)$ in a general basis, with no reason for any relation between the $(e h^\dagger (D^2)^{n-1} h e^c)$ and the $(e (D^2)^n e^c)$ operators. It is the total derivative structure which guarantees that even with arbitrarily complicated loop interactions involving general masses and  Yukawa and scalar interactions of the $\Psi_I, \sigma_a$, etc., the matrices ${\bf B},{\bf C},..$ and ${\bf \tilde{B}}, {\bf \tilde{C}},..$ are equal up to numerical factors. This implies an infinite number of relations between higher dimension operators that are completely insensitive to the details of UV physics.

We have given a rather technical,``off-shell" understanding for the vanishing of the leading contribution to $(g-2)_\mu$ in our model, it would be nice to find a more conceptual explanation. For instance, there should be a simple ``on-shell" understanding of the phenomenon. After all, the leading left-right operator contributing to $(g-2)/m$ is most invariantly thought of as the coefficient of three-particle amplitude for two {\it massless} fermions and a photon, with all $+$ or all $-$ helicity, as ${\cal A}(1+2+3+) =[(g-2)/m] [13][23]$. There must be a simple on-shell understanding for why this massless three-particle amplitude is not forced on us at one-loop, but can be generated at higher loop orders or subleading orders in $(m_W^2/M^2)$. 

It would also be interesting to find a pure symmetry argument that makes the result completely obvious without refering to any analysis of diagrams. Ordinarily when symmetries protect the generation of operators at loop level, the reason is seen scale-by-scale, while in our example, the loop integrand does not vanish and there is no scale-by-scale understanding of the zero. Thus any symmetry explanation must be ``not Wilsonian" in the sense of having to explain cancellation between widely distant scales. Note that in our simplest example, there is a precise cancellation between the contributions from around the scale $m_S$ and $m_L$, and so one might be tempted to look for a symmetry exchanging these scales. However, as we have mentioned,  the same zero occurs with any number of insertions on the fermion line, where $f(u) \to \frac{u}{\prod_j (u + m_j^2)}$. In this case partial fractioning tells us that the contribution from the scale $m_j^2$ is proportional to $\prod_{i \neq j} \frac{1}{m_j^2 - m_i^2}$, and there is no discrete symmetry relating the contributions at the different scales to each other; it is simply that the full sum vanishes. Recall also that as discussed just above, the vanishing of $(g-2)$ is not just a one-loop statement, but extends to any interactions at higher loops that only dress the neutral fermion line in the diagram, so any putative symmetry must also explain this fact. 

The ``integrand is a total derivative" phenomenon of this note has been seen in other settings. A concrete example in the context of a special non-supersymmetric string compactification was studied long ago in~\cite{Moore:1987ue}, where as a consequence of a hidden ``Atkin-Lehner" symmetry--not part of the usual modular group--the one-loop vacuum energy in a particular two-dimensional, non-supersymmetric compactification of string theory was seen to vanish. This was soon thereafter understood as a total derivative phenomenon~\cite{Lerche:1987qk}: the integral over the fundamental domain is a total derivative, whose boundary contributions happen to vanish in the model. 

More recently, the total derivative phenomenon (amongst other things) has been discussed in the context of toy integrals illustrating how ``UV-IR corrleations" might be relevant for addressing naturalness puzzles~\cite{KITP}. The idea can be illustrated by a toy integral. A typical loop integral for the Higgs mass looks like $\delta m^2 = \int dk^2 = \int_0^\infty du$ which exhibits the usual quadratic divergence. Let's suppose the theory is modified in the UV so the integrand becomes instead $\int_0^\infty du \frac{1}{(1 + \alpha u)^2} = \frac{1}{\alpha}$. Clearly we have introduced ``new physics" at the scale $1/\alpha$ to make the integral calculable/finite, and the result is set by that scale. The usual logic of naturalness would suggest that if we want to modify this result, in order to suppress the contribution significantly relative to $1/\alpha$, we would have to modify the integrand/introduce ``new physics" around the scale $1/\alpha$. 

But here is a simple counterexample, where the integrand is modified only at scales arbitrarily far above $1/\alpha$, but which makes the integral vanish. Consider the deformation 
\begin{equation}
\delta m^2 = \int_0^\infty du \frac{1 - \epsilon u^2}{(1 + \alpha u + \epsilon u^2)^2} = \begin{array}{ll} \frac{1}{\alpha} & {\rm for} \,\, \epsilon = 0 \\ 0 & {\rm for} \,\, {\rm any} \,  \epsilon > 0 \end{array}.
\end{equation}
This shows that at least in principle, ``new physics" at arbitrarily high scales $\sim 1/\sqrt{\epsilon}$ could cancel the low-energy contribution to $\delta m^2$ from around the scale $1/\alpha$.
Mechanically, this example was engineered as a total derivative: 
\begin{equation}
\delta m^2 = \int_0^\infty du \frac{d}{du} f(u), \, f(u) = \frac{u}{1 + \alpha u + \epsilon u^2}.
\end{equation}
We have $f(u \to 0) \to 0$, so the integral is given by $f(u \to \infty)$. When $\epsilon = 0$, $f(u \to \infty) \to \frac{1}{\alpha}$. But for {\it any} $\epsilon >0$, $f(u \to \infty) \to 0$ and the integral vanishes. This illustrate how some sort of UV/IR connection could be relevant in affecting an imagined computation for the Higgs mass.  Of course in this note we have seen the same phenomenon, where the total derivative arose automatically from the external momentum factors associated with the $(g-2)$ operator. But absent such a justification,  certainly in this toy example the invocation of the total derivative structure looks quite contrived. 

But there is perhaps a more interesting way of thinking about what we have done. Let us go back to our original integral, and this time think about it as a contour integral in the complex plane. It is more transparent to think of this as integrating a 1-form in ${\mathbb{P}}^1$. Concretely we use a co-ordinate $\lambda_i$ identified up to overall scaling as $\lambda_i \sim t \lambda_i$, so that in a particular co-ordinate patch we can use $\lambda_i = (1, z)$. We introduce the points $p_0 = (1,0)$ (``at zero")and $p_{\infty} = (0,1)$ (``at infinity"), as well as a point $q_{\alpha}=(-\alpha,1)$. Now consider a one-form $\Omega = \frac{\langle \lambda d \lambda \rangle}{\langle \lambda q_\alpha \rangle^2}$, (here brackets denote contraction with $\epsilon^{ij}$). We will first integrate $\Omega$ on an open contour between $p_0$ and $p_\infty$: $\delta m^2 = \int_{p_0}^{p_\infty} \Omega$. Putting $\lambda = p_0 + u p_\infty$, we have $\Omega = \frac{du}{(1 + \alpha u)^2}$ and the contour can be taken to run from $u=0 \to \infty$. Of course we expect that the integral of this one-form on an open interval will give something non-zero, and it does.

But let us now modify the story, not by changing the form $\Omega$, but by {\it modifying the contour of integration}. We will now take a {\it closed} contour, where $\lambda$ begins and ends at $p_0$, but which hangs around close to $p_{\infty}$ for a good stretch. To whit, we put $\lambda=p_0 + u p_\infty + \epsilon u^2 p_0$. When $\epsilon$ is small, the region where $u$ is small is clearly ``the IR" and the region where $u$ is large but $\epsilon u^2$ is still small is the ``UV", where $\lambda$ is close to the point $p_{\infty}$ at infinity. But eventually for {\it ultralarge} $u$, with $\epsilon u^2 \gg 1$, this ``deepest UV" region returns to the ``IR" again, and $\lambda$ returns to $p_0$ on the contour, so the contour is closed. (To be explicit, note that at large $u$, we have $\lambda \to \epsilon u^2 p_0$, which is projectively equivalent to $p_0$; this is the reason for our seemingly over-fancy thinking in terms of $\mathbb{P}^1$, to  make it clear that we are dealing with a closed contour here). 

Now on this contour, $\Omega = \frac{du (1 - \epsilon u^2)}{(1 + \alpha u + \epsilon u^2)^2}$, which is just what we found above integrates to zero. In our new interpretation this happens for a trivial topological reason: the form $\Omega$ with a simple double pole is exact, and we are integrating it over a closed contour, so it vanishes due to Stokes theorem. But note the ``zero" arises as a cancellation between two natural regions of integration.  The first part of the integration where $u$ goes from $0$ to something very large but for which $\epsilon u^2$ is not large,  which in a loose analogy might be thought of as the part of the computation with a ``field theoretic interpretation" and where the integrand is close to our starting point ``field theoretic" expression, is being cancelled precisely by the second ``deep UV region", as a simple consequence of the contour being closed, or having ``the deep UV being equivalent to the IR". 

This is trivial toy example, but at least suggests a concrete fantasy of a way in which UV-IR connections might be of relevance for challenging conventional notions of naturalness, in a way that would show up mechanically as a ``total derivative phenomenon" (which here is the direct, low-brow way of seeing the zero for the integral guaranteed by Stokes theorem). 

Mercifully returning to earth from these fanciful flights of speculation, in this note we have encountered the ``total derivative phenomenon" in a simple physical computation, possibly relevant to the real world, albeit violating naturalness for higher-dimensional $(g-2)$ operators, rather than relevant operators associated with the Higgs mass and the cosmological constant. While this finding may well just be a curiosity, it encourages a renewed effort to look for an analogous mechanism that might be relevant for addressing these most dramatic apparent failures of naturalness of our times.

\section*{Acknowledgments} 
We thank Raman Sundrum, Lawrence Hall, Greg Moore, and Giancardo Rattiudice for comments. We thank Radovan Dermisek, Keith Hermanek, and Navin McGinnis for useful discussions on $(g-2)$ in general vector-like lepton models and for pointing out an error in our expression for $\delta g_R$ in v1. We thank Admir Greljo  for pointing out the strong constraints on lepton universality from $\tau$ decays. We also thank Anson Hook for enjoyable discussions and comments, which encouraged us to further clarify the total derivative phenomenon.  This work was supported in part by the DOE grant DE-SC0009988 (N.A-H) and Friends of the Institute for Advanced Study (K.H.).

\bibliography{reference}

\end{document}